# Hector - a new massively multiplexed IFS instrument for the Anglo-Australian Telescope

Julia J. Bryant[1a,b,c], Joss Bland-Hawthorn[b,c], Jon Lawrence[a], Scott Croom[b,c], David Brown[a], Sudharshan Venkatesan[a], Peter R. Gillingham[a], Ross Zhelem[a], Robert Content[a], Will Saunders[a], Nicholas F. Staszak[a], Jesse van de Sande[b], Warrick Couch[a], Sergio Leon-Saval[b], Julia Tims[a], Richard McDermid[d], Adam Schaefer[a,b,c]

[a]Australian Astronomical Observatory, North Ryde, NSW 2113, Australia;
[b]School of Physics, The University of Sydney, NSW 2006, Australia;
[c]ARC Centre of Excellence for All-sky Astrophysics (CAASTRO)
[d]Department of Physics and Astronomy, Macquarie University, NSW 2109, Australia

## ABSTRACT

Hector[1,2,3] will be the new massively-multiplexed integral field spectroscopy (IFS) instrument for the Anglo-Australian Telescope (AAT) in Australia and the next main dark-time instrument for the observatory. Based on the success of the SAMI instrument, which is undertaking a 3400-galaxy survey, the integral field unit (IFU) imaging fibre bundle (hexabundle) technology under-pinning SAMI is being improved to a new innovative design for Hector. The distribution of hexabundle angular sizes is matched to the galaxy survey properties in order to image 90% of galaxies out to 2 effective radii. 50-100 of these IFU imaging bundles will be positioned by 'starbug' robots across a new 3-degree field corrector top end to be purpose-built for the AAT. Many thousand fibres will then be fed into new replicable spectrographs. Fundamentally new science will be achieved compared to existing instruments due to Hector's wider field of view (3 degrees), high positioning efficiency using starbugs, higher spectroscopic resolution (R=3000-5500 from 3727-7761Å, with a possible redder extension later) and large IFUs (up to 30 arcsec diameter with 61-217 fibre cores). A 100,000 galaxy IFS survey with Hector will decrypt how the accretion and merger history and large-scale environment made every galaxy different in its morphology and star formation history. The high resolution, particularly in the blue, will make Hector the only instrument to be able to measure higher-order kinematics for galaxies down to much lower velocity dispersion than in current large IFS galaxy surveys, opening up a wealth of new nearby galaxy science.

## 1. INTRODUCTION

Galaxy evolution is the grandest of all environmental sciences. Just how a galaxy forms and evolves in a given environment is one of the great unanswered questions in astrophysics. Hector is a revolutionary new AAT facility being built by the Australian Astronomical Observatory (AAO) and the Sydney Astrophotonic Instrumentation Laboratory (SAIL) at the University of Sydney. It will obtain integral field spectroscopy for 100,000 galaxies at z ~ 0.1 over ~3000 square degrees. While Hector builds on the success of the current Sydney-AAO Multi-object Integral field spectrograph (SAMI) instrument[4,5], it will enable fundamentally new science that cannot be undertaken with existing or planned instruments. Compared to SAMI, Hector will have a wider field of view (3 degrees vs. 1 degree), higher spectroscopic resolution (R=3000-5500 vs. R=1700-4500) and many more IFUs (50-100 vs. 13).

We present the science motivation for Hector in Section 2, followed by a description of the instrument design in Section 3. The IFU imaging fibre bundles, Hexabundles, described in 3.1 will be positioned by robots called Starbugs, described in 3.2. Fibres from the hexabundles will then feed into the spectrographs discussed in Section 3.3. Hector requires a new 3-degree-field top end for the AAT for which the mechanical and optical design tests are shown in Section 3.4.

---

[1] For correspondence: jbryant@physics.usyd.edu.au

# 2. SCIENCE MOTIVATION

The overarching science goal is to understand the physical basis for the diversity of galaxy properties in the local Universe. Substantial steps forward have been possible with smaller surveys such as SAMI, which have sufficient sample sizes to study the properties of galaxies as a function of their mass and local environment, but the diversity of galaxies is driven by more than just these two parameters. The accretion and/or merger history of a galaxy is fundamental to its morphology and star formation history, but to date it has been hard or impossible to capture the diversity of accretion histories observationally. The broader large-scale environment is also known to be important in determining a galaxy's structure via tidal torques. Only with Hector targeting 100,000 galaxies can we hope to probe the detailed physics of galaxy formation in these two extra dimensions of accretion history and large-scale environment. Specifically, the integral field data from Hector can provide specific angular momentum, higher order velocity moments, mis-alignments between gas and stars as well as measure dynamical disturbance. All of these tracers provide a view into accretion and merger history, which when combined with the unprecedented sample size of Hector will allow us to connect a galaxy's specific history to its current state. By connecting these properties to the network of large-scale structure we will be able to show how accretion is modulated within the cosmic web.
*With Hector we will decipher the diversity of galaxies through understanding their individuality.*

Galaxies share important properties like halos, disks and bulges. But when we look in detail, no two galaxies are identical, and we can see substantial differences even between relatively featureless spheroidal galaxies. Many have tried to identify a surrogate or a twin for the Milky Way. For example, Efremov (2011)[6] identified the 8 best candidates for surrogates, all of which look visibly different. Some of these differences may reflect internal processes and different stages of evolution, but others may reflect variations in the environment. Interestingly, even the relatively low resolution of the best cosmological simulations today finds that essentially *all* galaxies are different[7]. The central questions are, ***what are the main drivers of galaxy diversity? And, how do they manifestly influence galaxy properties?*** Galaxy stellar mass is widely seen to be the primary driver, with strong trends of mass with star formation[8,9], metallicity[10], and morphology now well characterized. Beyond stellar mass, the next well recognized driver of galaxy diversity is environment. While there has been much debate concerning the best metric to characterize environment, it is now thought that the mass of the host dark matter halo is the primary environmental driver of galaxy properties. It should be noted, however, that measuring the mass of dark matter halos remains a challenge, particularly at low mass.

Even once stellar mass and dark matter halo mass are accounted for, there remains a huge diversity of galaxy structural properties and star formation histories. It is clear that more than just mass and small-scale environment are required to properly understand the pathways to galaxy growth. The well-established picture of hierarchical growth of galaxies tells us that the accretion history of a galaxy is fundamental to its evolution. For example, does a galaxy go through one or more major mergers, how many minor mergers occur and how much external gas is accreted? Unaccounted for, the stochastic nature of a galaxy's accretion history provides substantial variance in the population, obscuring the underlying physics.

It is also clear that the host halo mass cannot be the only environmental factor. Differences are already seen from small-scale environments between galaxies that are centrals or satellites in a halo. Outside the halo, the large-scale environment defined by the geometry and topology of structure, e.g. within walls, filaments or voids can also modulate the growth of galaxies via both tidal torques and gas accretion.
In summary, the dominant drivers of properties of galaxies are likely to be:
1. Galaxy stellar mass.
2. Host dark matter halo mass.
3. Accretion history.
4. Halo location (central vs. satellite).
5. Large-scale topological environment (walls, filaments, voids etc).

Next generation galaxy surveys need to have sufficient statistical precision to be able to meaningfully span the above parameter space, but more importantly, new breakthroughs will be possible if we have measureables that will enable us to unravel the accretion history of a galaxy. Crucially, integral field spectroscopy gives multiple windows into these important parameters, especially when combined with the latest generations of numerical galaxy simulations (e.g. [7]). Hector will allow us to measure
  i)  spin parameters and specific stellar angular momentum that traces differing merger histories;

ii) the higher-order kinematics of the line-of-sight velocity distribution that can differentiate between a variety of formation paths;
iii) measure current kinematic disturbance as a probe of ongoing dynamical interactions;
iv) measure the mis-alignment between gas and stellar kinematics to identify gas of external origin;
v) measure stellar and gas phase metallicity gradients as tracers of external (low metallicity) accretion.

On scales larger than a single halo, the global environment defined by large-scale topological structure (filaments, sheets, voids etc.) adds yet another level of complexity. The inflow of gas that fuels star formation is in competition with feedback processes and this inflow of gas is crucially dependent on the *details* of where the galaxy sits within the large-scale structure. Cold flows from filaments deliver the fuel for star formation, but the resulting star formation rate is in turn dependent on the angular momentum from the filament[11]. This in turn depends on the alignment of galaxy spins with large-scale structure that simulations predict can be either aligned with filaments[12] or at the highest halo masses remain perpendicular to the filament[13]. Observational results have shown vastly differing conclusions[14,15,16]. The requirement to make a breakthrough in this area is a large-scale survey of galaxies which can measure angular momentum and is carried out in a region which has sufficiently high sampling to provide well-defined large-scale structure. Theoretical work by Trowland et al. (2013)[12] and Dubois et al. (2014)[17] suggest 60,000 and 150,000 galaxies respectively are required to disentangle spins with large-scale structure.

The driving parameters of galaxy formation lead to observable differences in a range of dependent quantities that Hector will be uniquely positioned to measure. These include current star formation and star formation history, stellar metallicity and age gradients, gas phase metallicity, angular momentum, kinematic morphology and many others. Bringing all of these together is crucial to uncovering the true nature of galaxies. A prime example is shown in Figure 1. This example[18] from the SAMI Galaxy Survey is a galaxy in a high-density group environment that has centrally concentrated current star formation (as traced by Hα). This would appear to be an excellent candidate for outside-in quenching (for example from ram-pressure stripping), but the stellar populations of the outer disk are at least several Gyr old and not recently quenched. More telling yet is the kinematics that shows gross misalignment between gas and stellar motions in the outer regions, and the stellar metallicity is lower in the central young stars, associated with current star formation. Bring all this information together we find that this galaxy shows evidence for nuclear triggering of star formation due to external gas accretion, a completely different picture to simple outside-in quenching.

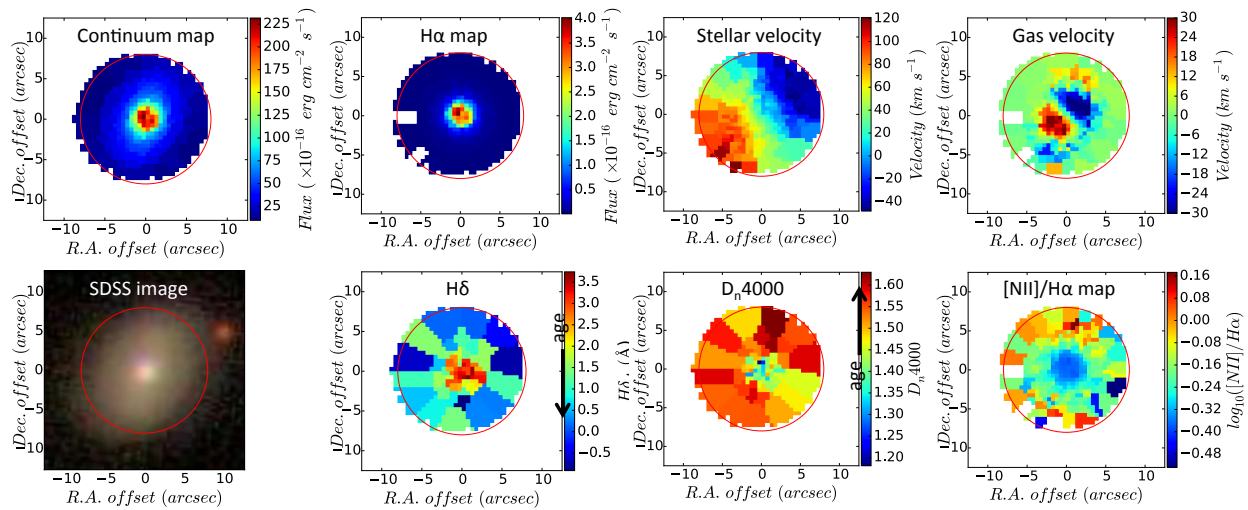

Figure 1: GAMA ID 517302, a highly unique galaxy. SAMI continuum and Hα images (top left & centre left), SAMI stellar and gas velocity (top centre right & right). The star formation is concentrated in the galaxy centre compared to the extent of the continuum, and the gas and stellar velocities are highly misaligned. SDSS image is lower left. Lower centre left and right show the radial distribution of effective ages measured by emission-subtracted Hδ absorption equivalent width index and the $D_n4000$, showing that the stars at higher radius are older. [NII]/Hα ratio map (lower right) shows elevated [NII]/Hα toward the edge of the galaxy which may indicate shocks or a post-AGB stellar population.

The Hector observations need to be paired up with detailed analysis of the local and extended environment of each galaxy, and this can be provided by next generation galaxy surveys such as TAIPAN[19] and WAVES[20]. Coupled with this, the surveys for HI gas using ASKAP[21] supply the missing neutral gas that cannot be detected in the optical. Hector, ASKAP and TAIPAN/WAVES then provide a fantastic synergy that will allow us to delve into the details of a galaxy's formation history like never before.

Hector will be the first spatially-resolved integral field spectroscopy (IFS) survey with the >60,000 galaxies necessary to span the parameter space described above and connect its physical characteristics to its stellar mass, small-scale environment, large-scale environment and accretion history. These new "measurables" will decode each galaxy's genome that defines its individuality and will then unpick the strongest impacts on the history of that galaxy's formation. The new innovation offered by Hector is the combination of a huge galaxy sample with the rich diversity of information enabled by high-resolution integral field spectroscopy. These are both necessary to understand the complex interaction of mass, environment and accretion history on angular momentum, structure formation and stellar mass growth. Never has there been a large enough observational sample to apply these techniques and then directly compare to the simulations. Hector will be the only instrument capable of reaching the galaxy numbers required in the near future because of the survey efficiency afforded by the unique combination of the 3 degree field and a telescope on which major surveys are prioritised, such that a very large program is possible.

## 3. INSTRUMENT DESIGN

Hector will consist of fibre imaging bundles, called hexabundles, that will be robotically positioned by autonomous fibre positioning robots, called 'starbugs', on the focal plane of a new 3-degree-diameter field (3dF) top end corrector. The ~40m fibre cable from the top end will feed new fixed format spectrographs stably mounted at dome floor level. Here we discuss the hexabundles, starbugs, spectrograph requirements and the mechanical and optical designs for the 3dF top end.

### 3.1 Hexabundles – fibre IFUs

Following the success of fused fibre IFUs with high fill-fraction, called hexabundles[22,23,24,25], new format hexabundles are being developed at the Sydney Astrophotonic Instrumentation Laboratory (SAIL) at the University of Sydney. Hexabundles have the buffer removed and cladding etched down over a short length. The fibres are then fused together in such a way as to have no more focal ratio degradation (FRD) than the original bare fibre. The resulting imaging bundles therefore have excellent optical performance with the advantage of a very high (~75%) fill fraction (active core area/total bundle area). The full details of the FRD, cross-talk and throughput performance from fused hexabundles is given in Bryant et al. (2014)[25]. Figure 2 shows an existing SAMI hexabundle in which the cores are not hexagonally packed. The new Hector hexabundles will have hexagonally packed cores of all the same size. The advantage of pursuing the hexabundle design is that they have proven to be robust with no degradation or breakages in the 13 SAMI hexabundles after 3 years of use. The main new development however is the mounting of the bundles in starbugs and testing is underway to ensure they remain as robust within a more compact housing that fits the starbug space envelope.

Hector's hexabundles will have a range in angular diameter of 15-30 arcsec per hexabundle. Each fibre will be ~1.6 arcsec diameter, leading to 61 to 217 fibres per hexabundle for 100μm-core fibres. The distribution of hexabundle sizes is based on detailed modeling of galaxies in the simulated Hector Galaxy Survey, and the science requirement that 90% of galaxies are imaged to two effective radii ($R_e$).

Extension of the hexabundles to a diameter matching $2R_e$ will allow additional science, beyond the SAMI instrument, on merging, environmental impact, and the dynamical, enrichment and star formation histories of galaxies. Competitive instruments such as CALIFA[26] and MANGA[27] partly probe to higher $R_e$, but will not have the sample size of Hector. Moreover, an increased radius increases the number of independent spatial elements across a hexabundle.
These larger hexabundles in Hector will also bridge the current gap between the SAMI field-of-view and the spatial resolution of upcoming ASKAP HI surveys because a 30" diameter bundle would be a better match to the ASKAP resolution. HI would then complete the picture beyond the Hector radius.

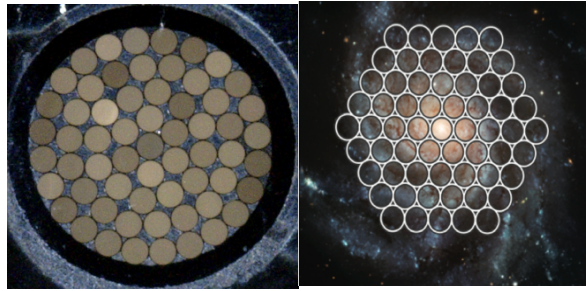

Figure 2: Hexabundles: The ~1mm diameter face of one of the current hexabundles on the SAMI instrument showing the 61 fibre cores (left) and the layout for the new hexagonally-packed design for Hector shown with 61 cores (right). Hector bundles will have up to 217 cores, which is the addition of 2 more hexagonal annuli to the picture shown.

## 3.2 Positioner: Starbugs

The 50-100 hexabundles plus sky fibres on Hector will be positioned on a glass field plate by robots called starbugs[28,29,30,31,32]. The Hector starbugs will be a progression of the technology developed to produce the 180 starbugs for the Taipan multi-object spectroscopic galaxy survey[16]. A starbug consists of two coaxial piezo-electric tubes that are electronically controlled to flex in a walking action in either of four directions or to rotate on its own axis as illustrated in Figure 4. This walking ability enables a starbug to position its payload within microns of a target. The starbugs walk on a glass field plate, which is positioned so the focal plane of the telescope is at the face of the science fibre bundle inside the starbug. The starbugs attach to the field plate by vacuum. The vacuum, or negative pressure within the starbug body which is routed from an external vacuum system, needs to be sufficient to overcome the mass and moments imparted by the payload (for Hector, this payload is the hexabundle). The Hector starbug will undergo design development for the increased holding force required compared to the existing TAIPAN technology. An increased holding force is needed both because the payload is heavier and the fibre bundle is more ridged. The holding force is proportional to the cross sectional area of the outer tube and the pressure differential between the atmospheric pressure and vacuum supplied. A prototype starbug with a 12 mm diameter will achieve twice the holding force compared to TAIPAN. This diameter will be increased or decreased as required during prototyping to ensure sufficient holding force.

Precise closed loop control of directional and rotational motion is enabled by an on-axis metrology camera and three backlit metrology fibres (fiducials) shown in Figure 3. The metrology camera needs to be facing the starbugs with consideration for focal length, CCD pixel size, starbug metrology fibre diameter, positioning precision requirement, light path shadowing cost and telescope geometry. Starbugs simultaneously proceed to their individual targets and the entire field can typically be configured within a few minutes, during the slewing time of the telescope. Fine positioning is then achieved while the telescope is on target by referencing guide bundles on guide stars to circumferential fiducials on the glass field plate. The position of the payload axis is characterised with reference to the three metrology fibres in each starbug so the payload axis is then positioned onto its target galaxy and rotated with reference to the circumferential fiducials on the field plate.

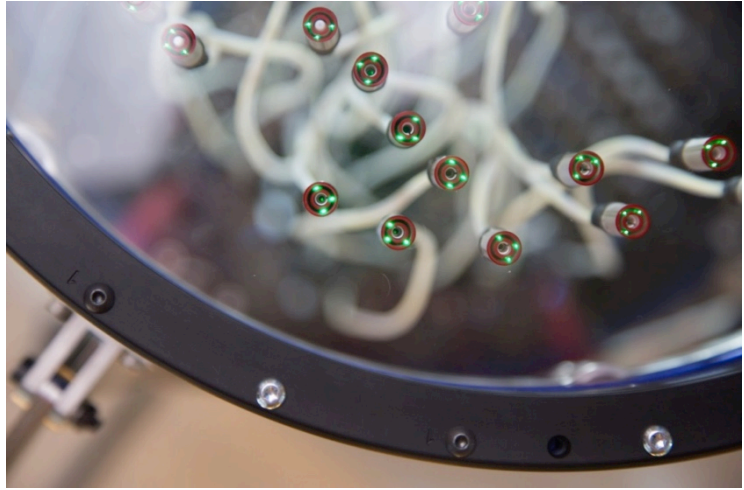

Figure 3: Starbugs showing the 3 illuminated metrology fibres

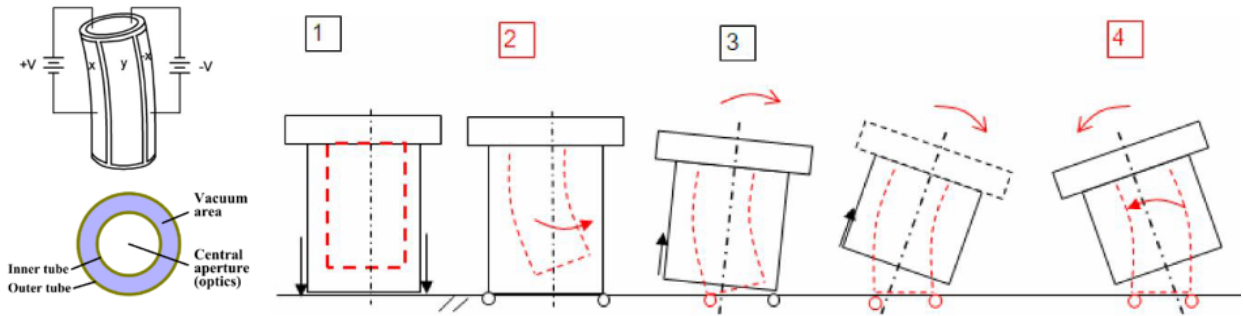

Figure 4: Starbug walking motion.

A starbug is a line replaceable unit to ensure there is no loss of efficiency if one component breaks. The TAIPAN starbug in Figure 5 has an internal optical fibre connection within the electrical plug. A polished fibre ferrule is fitted into the starbug using a precision push fit and therefore can be repolished or the entire payload can be replaced.

If the payload dock holding the ferule can be positioned outside the starbug then potentially the hexabundles may be able to be positioned closer together. Experiments will be conducted to investigate the feasibility of cantilevering the payload dock outside beside the starbug.

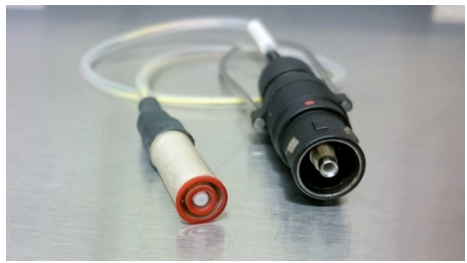

Figure 5 A starbug is a line replaceable unit. The plug on the right contains the optical fibre and vacuum connections to the starbug unit (left). The foot that contacts the glass field plate is seen in red, with the white polished fibre ferule in the centre.

A starbug instrument requires a movable Connector Plate, a Bug Catcher, a metrology camera and starbug drive electronics. The starbug plug connects to a movable Connector Plate that is moved toward the Glass Field Plate to increase the patrol radius of the starbug by allowing more slack in the cable. The Bug Catcher shown in Figure 6 is a structure that moves along the axis of the instrument between the connector plate and the glass field plate. It supports and protects starbugs while the telescope is moved when there is no vacuum. The bug catcher doubles as a positioning device to place starbugs against the glass field plate for vacuum adherence.

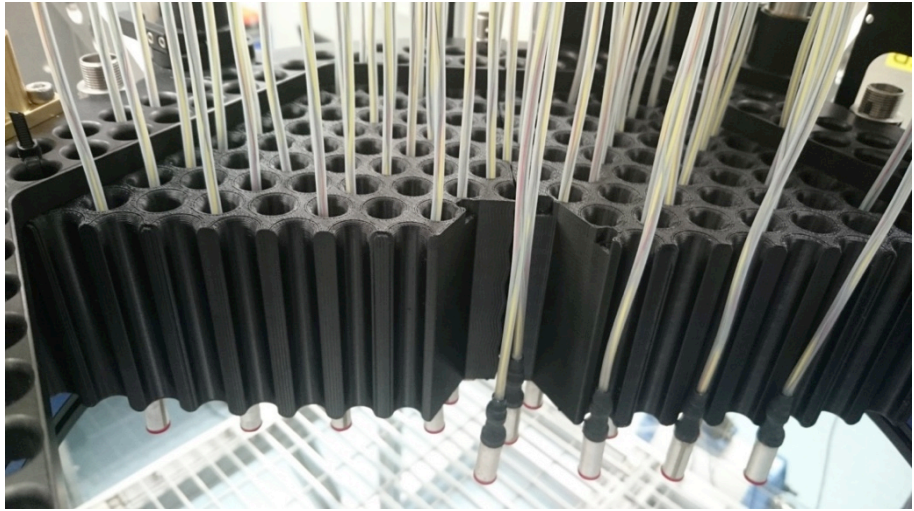

Figure 6: The starbugs are contained in the Bug Catcher that is used to position the starbugs when the vacuum is applied. Half of the Bug Catcher is removed here to show the protective tubes.

### 3.3 Spectrograph and requirements

The Hector spectrographs are designed to be modular and replicable. Several designs are currently being evaluated to meet the following science-based requirements.

(a) Wavelength Coverage

Figure 7 shows the essential emission and absorption lines required for Hector science. The Hector Galaxy Survey will extend to $z=0.15$ which sets the wavelength coverage required. There is an aim for the spectrograph to have continuous wavelength coverage from 372.7-776.1 nm, with a possible red arm upgrade to ~1μm.

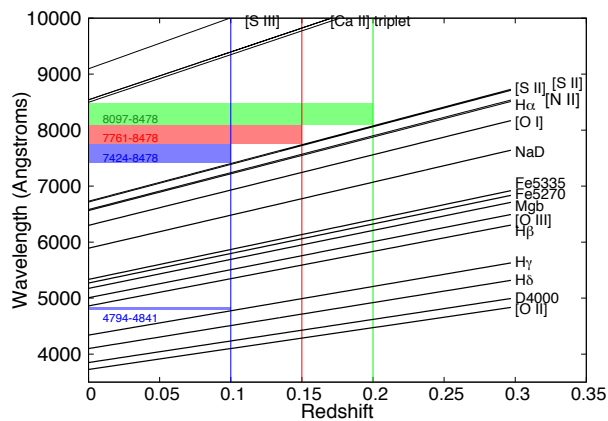

Figure 7: Wavelength coverage versus redshift for key emission and absorption lines. The coloured stripes with wavelength ranges in Å mark the wavelength regions (allowing a +/-20Å window) not required to be covered in order to observe the lines shown up to redshifts of 0.1 (blue), 0.15 (red) and 0.2 (green).

(b) Resolution and Sampling

The tightest constraints on the spectral resolution come from stellar kinematics science cases in order to measure crucial higher order moments ($h_3$ and $h_4$)[33,34] and the emission line science cases that require accurate de-blending of lines. Simulations of the stellar kinematics and emission line science line fitting have been done to optimise the resolution required. Figure 8 shows a simulation of the measurement of stellar velocity dispersion ($\sigma$) along with higher order moments $h_3$ and $h_4$. The aim is to measure these kinematic tracers in galaxies that have the lowest intrinsic dispersion ($\sigma_{in}$). Van de Sande et al. (2016)[35] has shown from SAMI Survey galaxies that low velocity dispersions are found in high-mass galaxies, not just in low-mass galaxies, and therefore low velocity dispersions need to be measurable across the full stellar mass range. A resolution of 1.3Å or better is required to measure stellar kinematics in galaxies that have $\sigma_{in}>$40km/s for S/N=10/Å whereas the resolution of SAMI (~2.75Å) limits the galaxies in which both $h_3$ and $h_4$ can be measured to those having $\sigma_{in}>$~130km/s (we note that at S/N=20/Å $\sigma_{in}>$~75km/s can currently be measured with SAMI and Hector will reach galaxies with $\sigma_{in}>$25km/s).

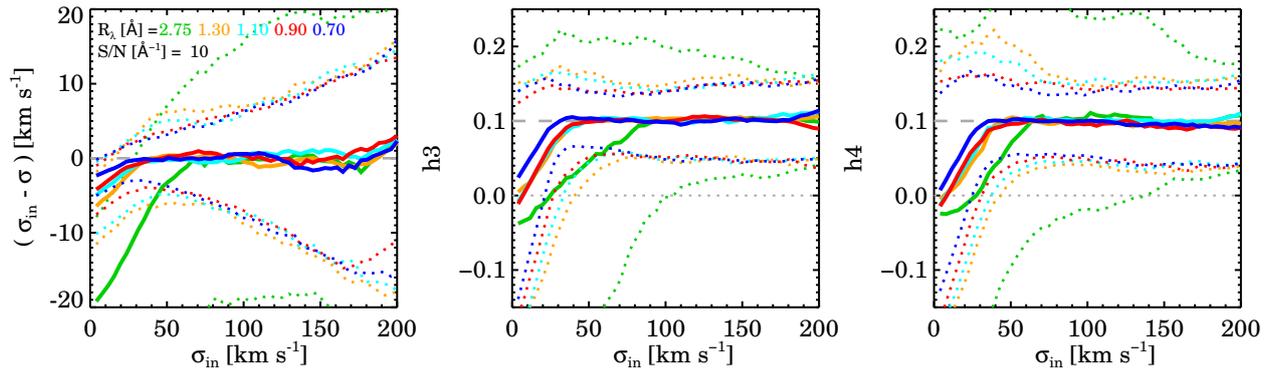

Figure 8: Impact of the resolution on the recovered kinematics for S/N = 10/Å. Different colours show different resolutions with R[Å]=2.75 matching the blue resolution of the SAMI instrument. In the left panel, the lines show the 50th (median, solid line), 16th and 84th percentiles (1σ errors, dotted lines) of the differences between the measured values and the input values of the velocity dispersion $\sigma_{in}$. The middle and right panel show the recovered values of $h_3$ and $h_4$, compared to the input values of 0.1 (dashed line). In order to measure $h_3$ and $h_4$, requires 1σ errors <0.1 (lower dotted line must stay above 0.0). Therefore for S/N=10/Å, a resolution of 1.3Å or better is required to measure stellar kinematics in galaxies that have $\sigma_{in}>$40km/s, whereas the resolution of SAMI (2.75Å) limits the galaxies in which $h_3$ and $h_4$ can be measured to having $\sigma_{in}>$~100 and ~130km/s respectively.

Accurate deblending of emission lines into components is crucial for:
1. Detecting winds and outflows in galaxies.
2. Separating star formation and AGN processes and identifying shocks with ionisation mapping.
3. Galaxy metallicities and chemical abundance mapping.
4. Kinemetry.
5. Scaling relations using gas rotations and dispersions.
6. Mapping star formation from H-alpha for comparison with environment, morphology, radio emission.
7. Kinematic misalignments and the origin of gas in galaxies.

Resolutions of R ($\lambda/\delta\lambda$)>3000 in the blue and R>5000 in the red are required to accurately constrain multi-component Gaussian fits. While higher resolution is generally beneficial for emission line science, increasing the resolution has to be traded off with additional read noise, particularly in the blue, plus increasing spectrograph cost and size.

The best balance between key stellar and emission line science cases sets Hector's spectral resolution at 1.3Å with 2 pixel sampling from 3727-7761Å, as this is the minimum for the emission line science requirements and higher resolution presents much smaller gains for continuum science.

3.3.1 Spectrograph Designs

Two main spectrograph designs are currently being optimised to meet these requirements and the final design will be selected by mid-2016:

(1) A fast (~F/1.35) transmissive camera and a single 4Kx4K detector shown in Figure 9. This design depends fundamentally on diamond-turned and Magneto-Rheological Finishing (MRF) post-polished $CaF_2$ aspheres, with a maximum aperture diameter ~200mm. The collimator is an off-axis F/3.26 Schmidt and the gratings have slanted fringes to avoid the transmission grating Littrow ghost. The fast cameras and slanted gratings mean that a resolution of 1.3Å is obtained with 100μm fibers for the core wavelength range 370-777nm. Total throughput of 70% or more will be achieved at the peak wavelength for each arm. Full details of this spectrograph concept can be found in Saunders et al. (2016)[36].

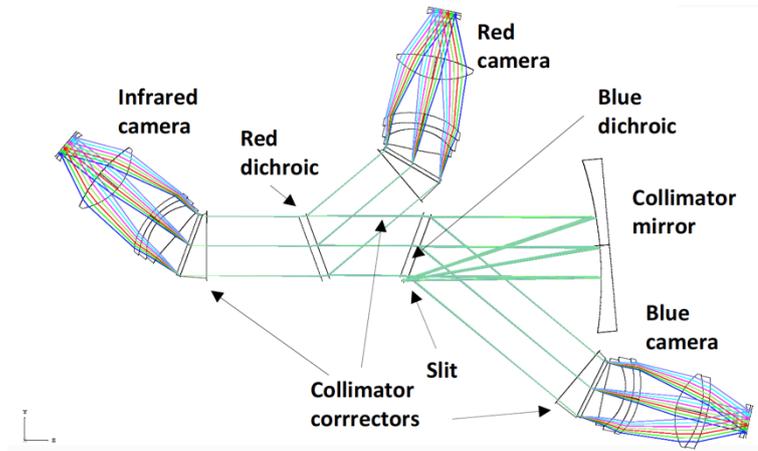

Figure 9: Overall layout for a 3-armed fast transmissive design for Hector. Note that a more compact layout is allowed by reflecting the blue beam to the other side of the collimator mirror, but this requires an increased red dichroic angle of incidence.

(2) A fully transmissive spectrograph relying on simple centered aspheres with small maximum P-V of 0.1mm and maximum slope of 0.015rad both with respect to the best fit sphere (Figure 10). Each camera has a 4K x 4K detector. All lenses have apertures smaller than 200 mm except two spherical lenses in the collimator that have apertures of 225 mm. The collimator is F/3.30 and the cameras F/1.40. An average resolution of 1.3Å is obtained in each camera with 100μm fibers over the full wavelength range of 372 to 778nm. An Infrared camera can be added by putting a dichroic after the third lens. There are 8 aspheres in the spectrograph with 3 on the lenses nearest to the focal planes of slit and cameras where they can have relaxed surface form tolerances. Full details of this spectrograph concept can be found in Content et al. (2016)[37]

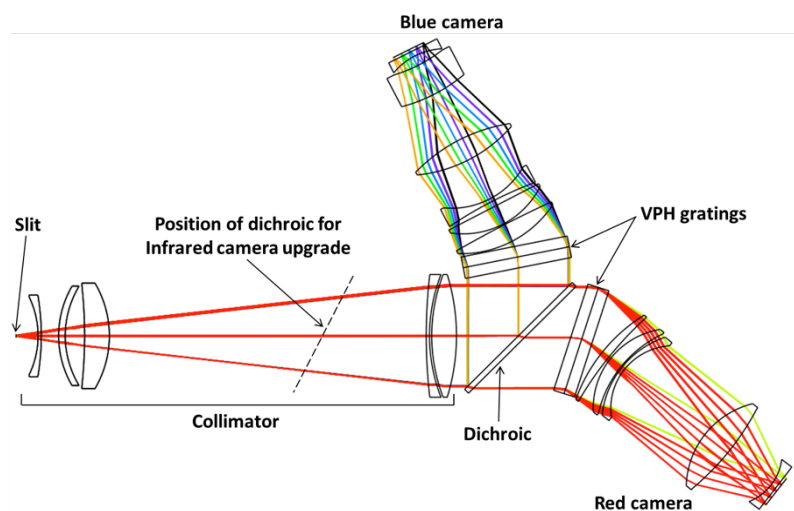

Figure 10: Layout of the spectrograph with simple aspheres. The 2 largest lenses of the collimator are 225 mm in diameter.

## 3.4 3-degree-field Corrector

### 3.4.1 General aims of top end design

There will be considerable scientific and operational advantages in providing a new top end with 3 degree diameter field (3dF) rather than sharing the 2dF top end with its existing spectrograph feeds or adapting it to suit the IFU-positioning robot. The new corrector can be mounted on an existing top end ring, retiring the original corrector with 1 degree field. To provide an alternative observing mode with minimal disturbance of the Hector mode, it is feasible to mount an f/8 secondary mirror for Cassegrain observing without disturbing the 3dF top end and its fibre feeds. Figure 11 shows how the secondary could be raised into position for attachment using a hoist on the existing overhead crane gantry.

The basic optical design for the wide field corrector (WFC), employing a novel means for atmospheric dispersion compensation (ADC), has been reported earlier[38,39]. It offers better imaging and higher throughput than the 2dF corrector as well as the obvious great increase in field area.

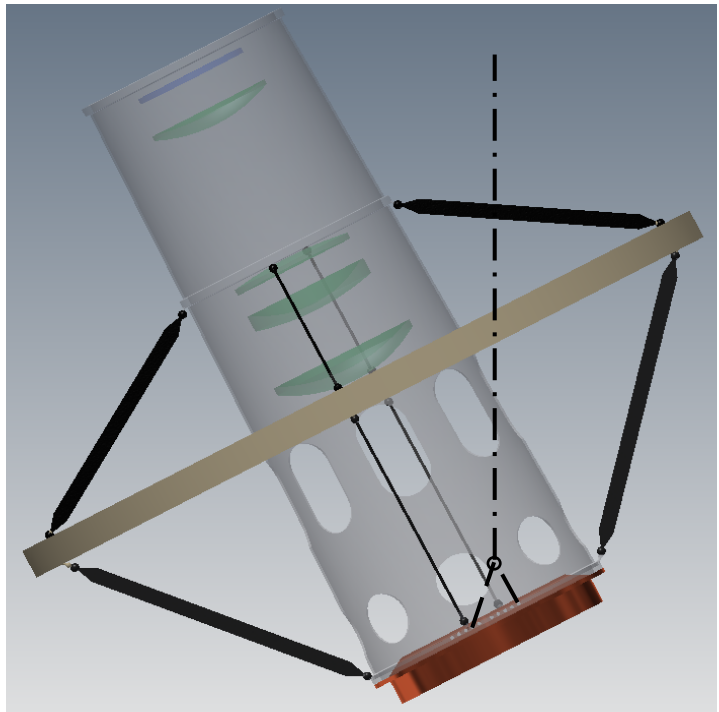

Figure 11: Elevation view of top end with telescope pointing ~36° due south of the zenith so that M2 can be raised from or lowered to the dome floor with the cables indicated by the dashed lines.

### 3.4.2 Atmospheric dispersion compensation (ADC) mechanism

For the ADC action, the second and the fourth of the elements are to be displaced laterally and tilted. With the AAT's polar axis mounting, this is complicated insofar as the directions of the offsets with respect to the corrector structure change with changes in the attitude of the telescope.

The lateral offsets for the two lens elements at a zenith angle of 60° are ~11 mm and 12 mm respectively while their tilts are ~ 0.34° and 1.23° respectively. These lens elements will be mounted in dual lens cells, with one movable cell for the ADC action. This cell will be supported against gravity by three part-spherical stainless steel inserts in the fixed cell with transfer ball type bearings linking the two cells and allowing motion of the inner cell in any lateral direction. Then the lateral position and the position angle of the element will be controlled using 3 encoded linear actuators for each ADC lens element. See Figure 12 for detail of the arrangement on element 4.

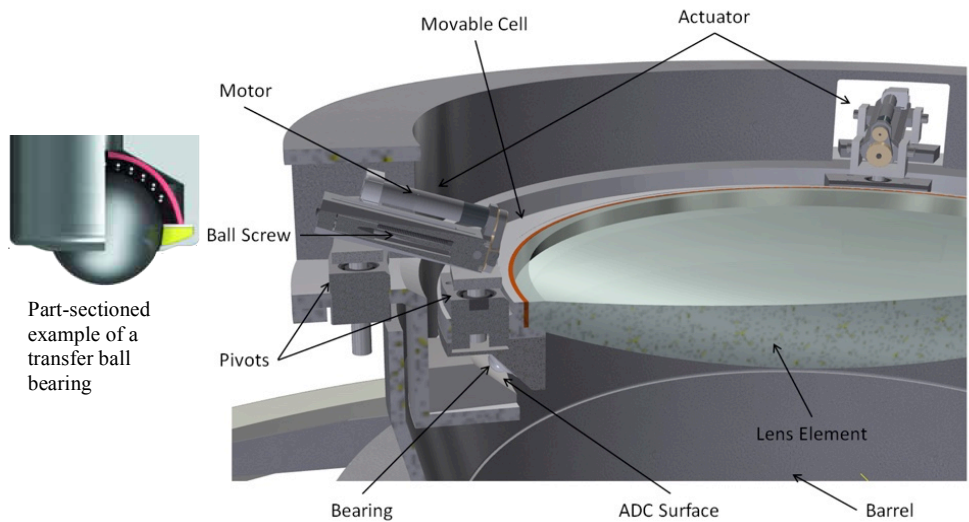

Figure 12: ADC mechanism for element 4.

### 3.4.3 Aluminum alloy for the corrector cells and main structure

Lens cells for large optics can be made of low expansion material that matches the expansion of the glass. Another option – an athermal design – is to combine inserts with a significant difference in thermal expansion, providing compensation. In the non-athermal approach, we use an aluminum barrel and lens cells. The impact of this choice on the optical performance is investigated.

Preliminary finite element analysis (FEA) in previous years indicated that it should be satisfactory to support the lens elements in a structure entirely of aluminum alloy, with substantial savings in weight and expense. More detailed FEA and optical analyses outlined here have confirmed that this is practical for 3dF.

Primary concerns in using an all-alloy design for a corrector are the stresses and deformations induced in the lens elements by thermal and gravitational loads and their effects on optical performance. The 3dF mounting design uses a non-athermal radial gap of 5mm and an intermediate RTV (room temperature vulcanization silicone) layer. This is sufficient to keep stresses low while maintaining the functional aspects of the lens elements The mountings of the lens elements to their respective cells use radial and axial RTV pads as were used for the Multiple Mirror Telescope[40]. This method is illustrated in Figure 13. The axial RTV pads will be pre-moulded and inserted into seats in the aluminum cell. The radial pads will be injected onto a foam mould during the installation of the lens elements.

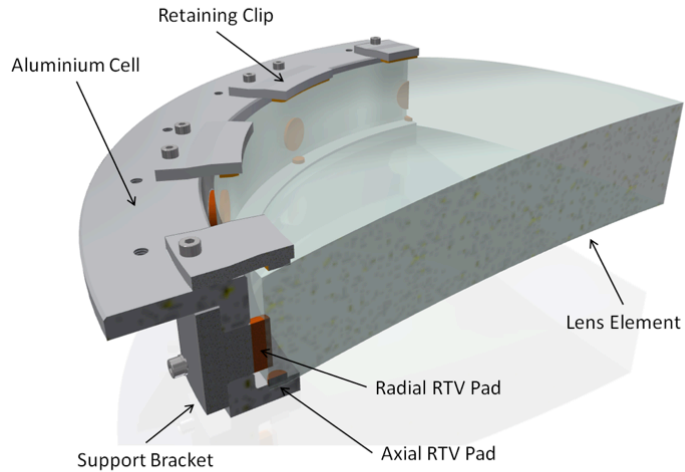

Figure 13: Lens Mounting Method

FEA has been performed using ANSYS on the 3dF corrector allowing for gravitational loading at the Zenith and at a Zenith angle of 45° and for a range of temperatures down to -4˚C. Results for surface deformations are shown for the most critical (front) element in Figure 14 and Figure 15.

The deformed shapes of the lens elements were exported from the FEA analysis to create new surfaces for optical analysis using Zemax ray tracing software. The displacements for the combined thermal and gravitational loads were calculated to be up to ~ 0.75 mm.. By far the majority of this was due to change in the axial spacing of the elements. The optical effect of this can be almost completely compensated by axial movement of the whole corrector, using the built-in AAT focus drive, which accurately moves the whole top end.

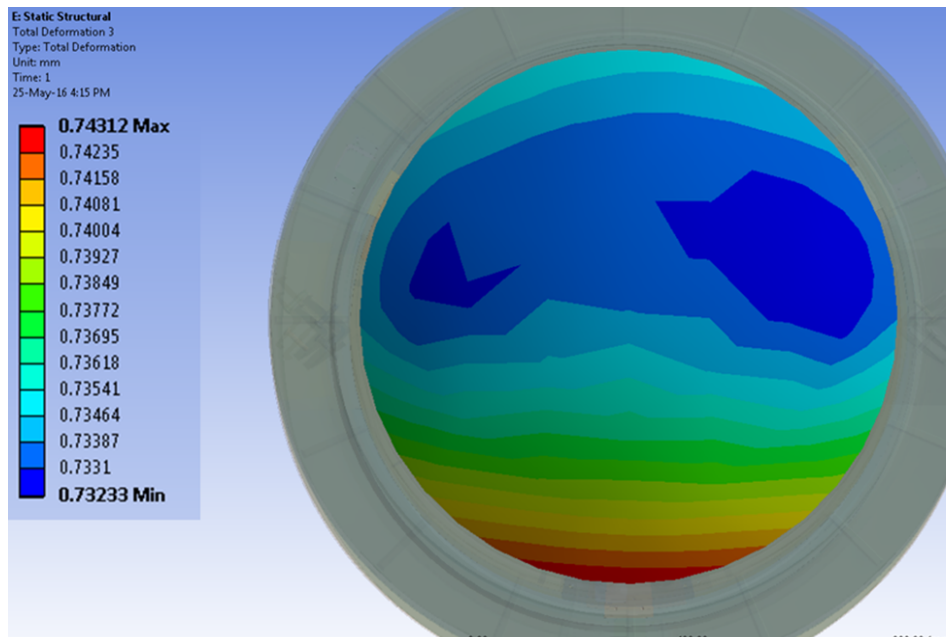

Figure 14: Deformations on the front surface of the front lens element of the 3dF corrector with temperature -2˚C and Zenith angle 45˚ relative to that with no gravity and temperature 22 ˚C. The contour range is 10.8 µm.

The corrector will be subjected to thermal differences after lab assembly and alignment at room temperature. The FEA has been performed for the temperature drop, ΔT= -24 deg C. The associated thermal effects are considered along and perpendicular to the optical axis of the corrector assembly.

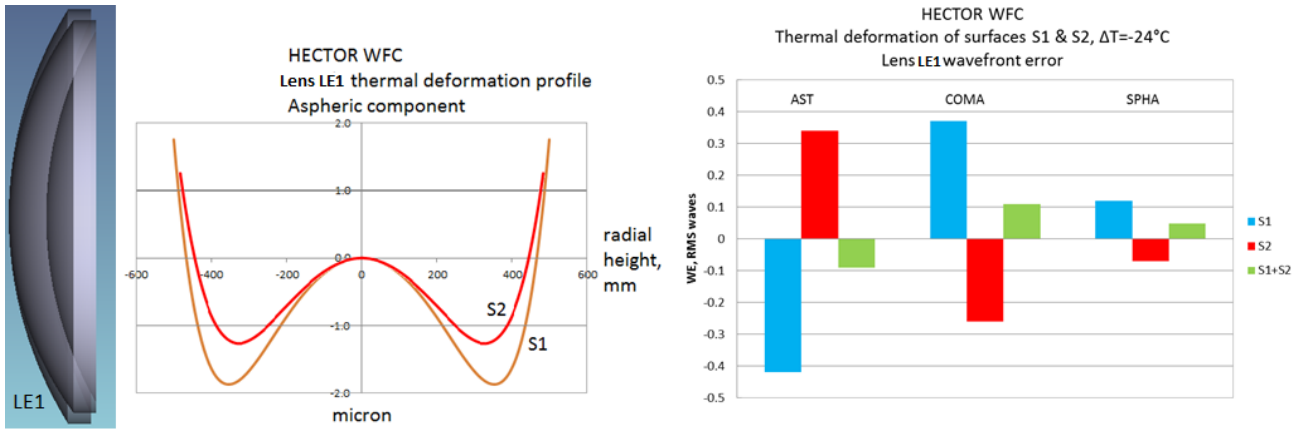

Figure 15: Element LE1 thermal and structural deformation.

Deformation modes induce asphericity on lens surfaces. Nominally spherical surfaces become aspheric upon thermal contraction. Best fit spheres are shorter upon contraction to yield ~30 μm sag difference at the edge. The aspheric profile of LE1 measures 3μm peak to valley (see Figure 15), and both surfaces acquire different asphericities due to slightly different radii and mounting arrangements. Aberration analysis confirms that the wavefront error contributions from both surfaces of LE1 nearly compensate each other, to 0.06 waves RMS, and the image quality remains unaffected. The residual design wavefront error is 0.72 waves RMS at 650 nm. The same conclusion applies to the element LE2, which exhibits an even higher asphericity of 6μm peak to valley. The effect of thermal deformation of elements LE3 and LE4 on the wavefront is an order of magnitude lower than for the lenses LE1 and LE2.

The analysed aspheric profiles include the structural deformation of the lens shape under gravity. The deformation under its own weight is most significant for the horizontal orientation of a lens. The net effect of gravity induced bending amounts to a maximum 2.5μm sag difference. This is small compared to the radii of curvature change. The radial load exerted by the shrinking lens cell dominates the thermal deformation profiles of the lens surfaces.

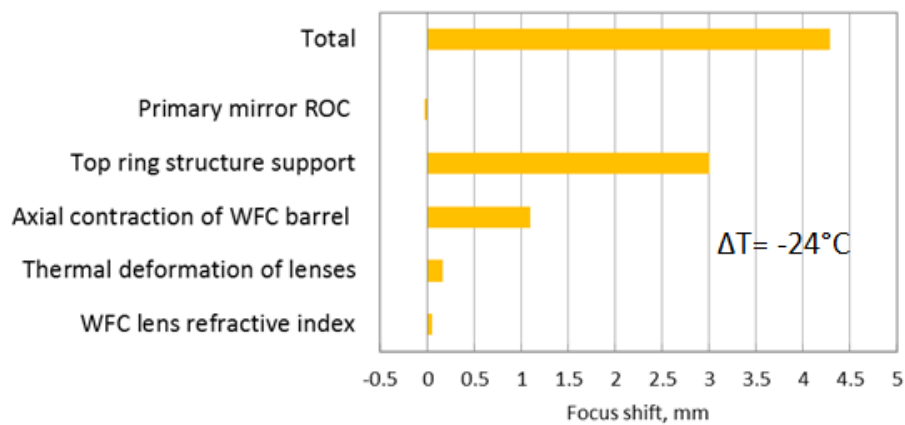

Figure 16: Thermal drift of the focal surface.

The aluminum barrel contracts along the optical axis resulting in the shift of all optical elements. Image quality analysis confirms that the element displacement affects only the location of the focal surface. The design image quality is restored after refocusing the telescope-corrector system. It was found that the aluminum barrel contraction over a 24 deg. C range will move the focus by 1.1 mm from the telescope primary mirror.

By comparison, the focal surface shift is influenced by a set of factors that are presented in Figure 16. The thermal dependence of the index of refraction of lens materials is taken into account. The thermal radial deformation of lenses results in the axial shift of each lens, as discussed above. Additionally, lenses are compressed, changing the optical power on each surface. All of these effects are much smaller than the thermal contraction of the corrector structure. The corrector is attached via vanes to the top end ring, which provides the interface to the steel telescope tube structure. Ultimately, the focal surface shift is dominated by the thermal variations of the length of this tube structure. The rate of the nominal focal surface thermal drift is estimated to be 0.18 mm per 1 degree C.

## 4. PREDICTED PERFORMANCE

In order to model the S/N requirements of Hector and predict the performance, a Hector Galaxy Survey simulator and a Hector total system simulation were developed. The Hector galaxy survey simulator incorporates the constraints of galaxy surface brightness distributions and galaxy sizes. In combination with the Hector Total System throughput simulation, we have assessed the predicted performance of Hector in terms of S/N for different galaxy types and the required hexabundles sizes to image the survey galaxies out to $2R_e$ with sufficient S/N.

The Hector Survey Simulator galaxy properties are shown in Figure 17. At any given redshift, the surface brightness of galaxies becomes fainter at lower stellar mass. This drives a galaxy selection in redshift and stellar mass that maintains the surface brightness being observed. The number of hexabundles across the 7.1 sq. arcsec (3 degree diameter) field needs to be less than the source density from the galaxy selection for efficient tiling of the field. A survey based on galaxies above the blue line gives a source density that allows two pointings per field and maintains both surface brightness and effective radius distributions that will reach required S/N and are large enough (in $R_e$) to give sufficient spatial elements across the galaxy. The galaxy selection above the red curve maintains the same surface brightness distribution but is less suitable as it includes many galaxies with much smaller $R_e$ (<~2 arcsec) which will not have sufficient spatial elements across the galaxy.

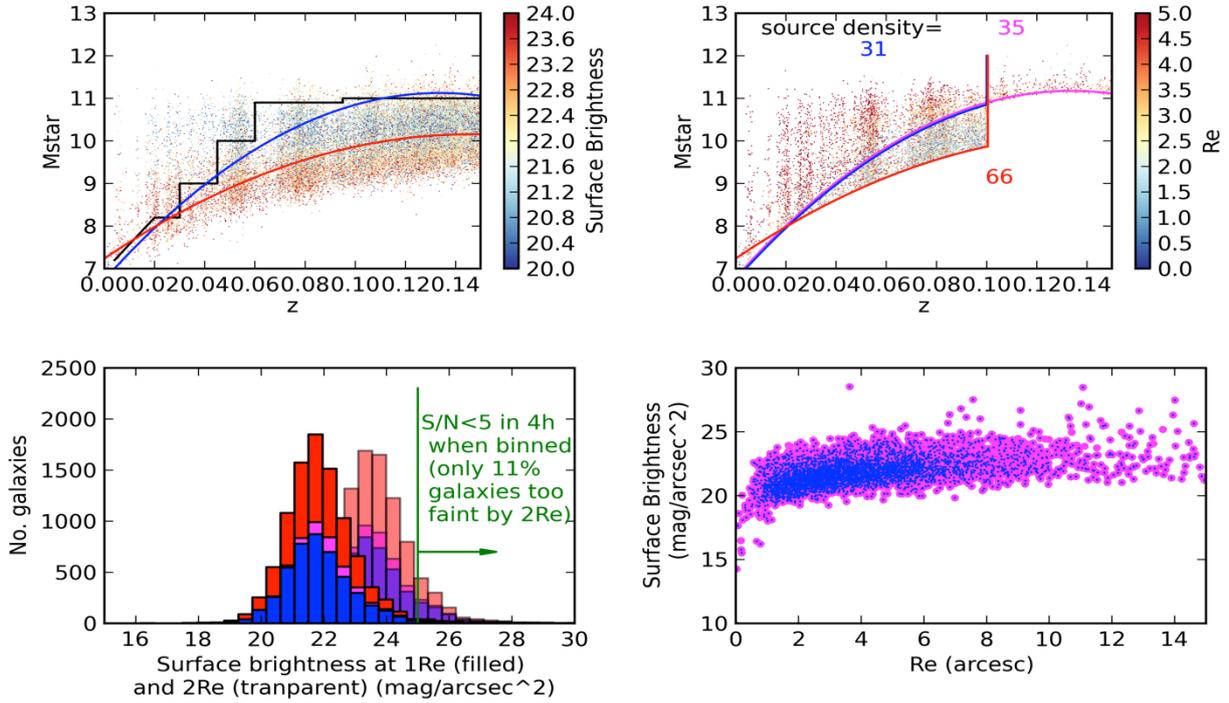

Figure 17: Using the Hector Survey Simulation code, simulations of several possible selections for a Hector galaxy survey based on the GAMA[41,42] catalogue (with objects flagged as having poor r-band fits removed as the surface brightness and $R_e$ used are in r-band) are shown. Top row: Redshift and stellar mass plane colour coded by surface brightness at $1R_e$ in mag/arcsec$^2$ (left) and effective radius ($R_e$, in arcsec) (right). The black line is the SAMI Galaxy Survey selection, blue is a fit to the lower stellar mass limits of that SAMI selection, red is a more extreme selection that attempts to optimise by surface brightness while still including a broad range in stellar mass. The blue lines and histograms in following panels are based on galaxies above the blue sample cut-off line, including only galaxies with z<0.1, while magenta are galaxies above the same selection cut-off but extended in redshift to z<0.15. The resultant source densities within 1 square degree on sky are given for each of the blue, magenta and red samples defined by galaxies that lie above their respective coloured lines. Second row: Distribution of r-band surface brightness at 1 and $2R_e$ (left). Note the red selection maintains the same distribution in surface brightness, as the SAMI-style selection in blue. R-band surface brightness at $1R_e$ vs $R_e$ (right) shows there is little dependence of surface brightness on $R_e$.

The Hector Total System simulator modelled throughputs including all components of the Hector system, from sky to detector. A spectrograph design with 1.3Å resolution, 100μm (1.6") aperture fibres was assumed. The resulting predicted throughput is shown in Figure 18.

A combination of the Galaxy and total system simulators, results in a wavelength dependent S/N for different galaxy types and profiles, that accounts for the r-band surface brightness at 1 and $2R_e$. Based on the simulated galaxy survey, galaxy colours and surface brightness distributions for models defined by Sersic index and stellar mass cuts were calculated. The S/N achievable in these galaxy classes was then measured and examples are shown in Figure 19, assuming 4 hours total integration comprised of 7 dithered pointings.

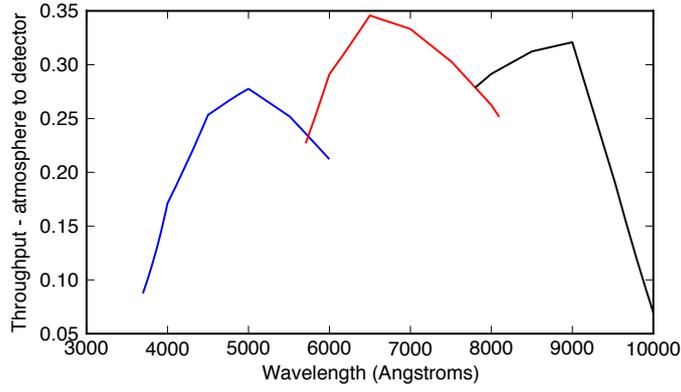

Figure 18: Predicted Hector throughput from sky to detector, based on the Hector total system simulator.

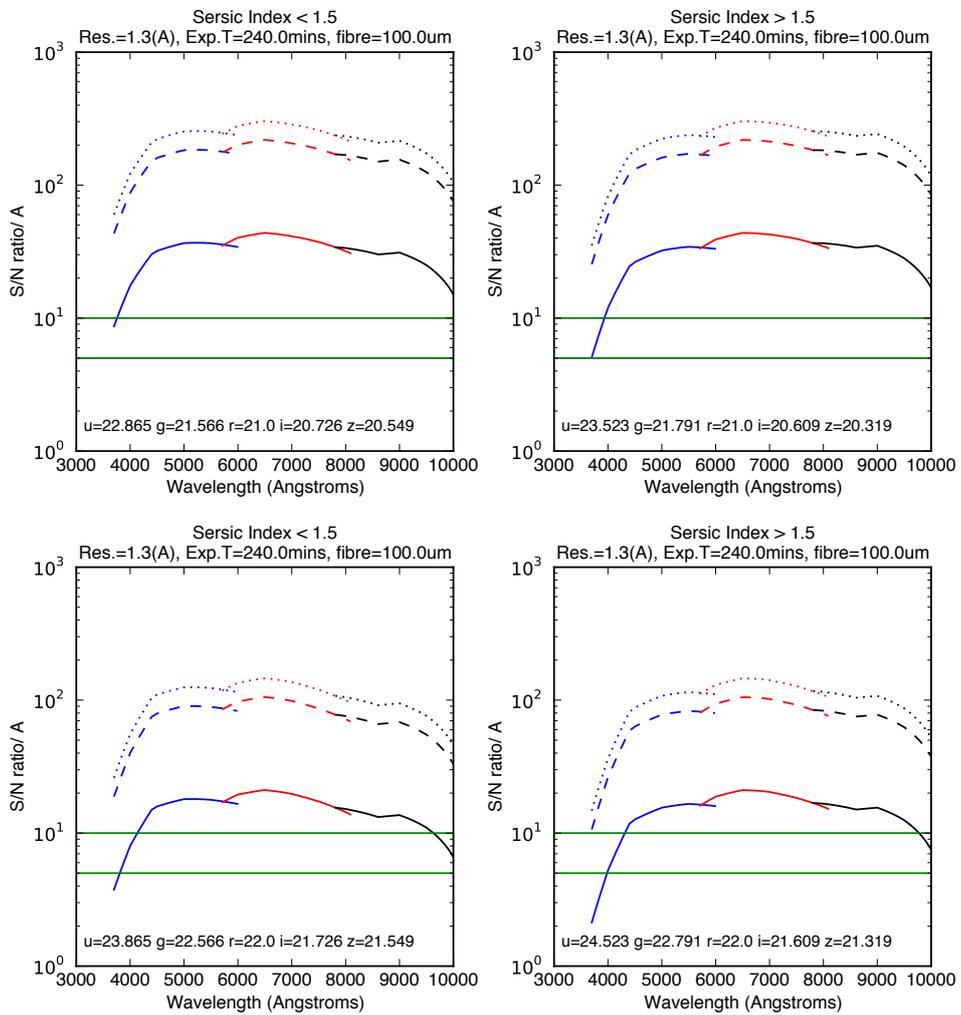

Figure 19: Signal-to-noise in 1.6" aperture (1 core) vs wavelength from the Hector total system simulation for 4 hours integration, 7 dithered pointings, and spectrograph design with 1.3Å resolution, 100μm fibres and a 1.6" aperture in dark sky conditions at Siding Spring Observatory, Australia. Galaxies were separated into Sersic index <1.5 (left column) and >1.5 (right column) which forms similar populations to that of a stellar mass cut around $10^{10}$ $M^*/M_{sol}$. The green lines mark a S/N

of 5 and 10/Å. The top and bottom rows model galaxies with r (SDSS AB) magnitude at 1Re of r=21.0 and 22.0 respectively. The modelled galaxy colours of each population have the surface brightness across each SDSS filter band as listed within each plot window. The solid lines are S/N in 1 fibre core, and the dashed and dotted are for 24 and 48 cores binned (i.e. the outer ring of 61-core and 217-core hexabundles) respectively.

The S/N requirements for Hector vary by the defined science cases. The stellar kinematics science requires continuum S/N>10 from 4700-7000Å but stellar populations requires continuum S/N>20/Å in each binned element. However emission line science cases need line S/N>5/Å out to 2Re. S/N requirements are therefore mostly constrained by the continuum S/N as emission line S/N is variable among galaxy types and difficult to predict. Based on the above simulations, Hector will achieve an extended object sensitivity of r(SDSS AB)=24.2mag/arcsec$^2$ in a 4 hour observation with SNR=10/Å from 470.0-700.0 nm and signal-to-noise ratio >2/Å at 372.7nm (for a galaxy with colours of g-r=0.79 at 1$R_e$) using a 7 dither-pattern observation. The simulated galaxy survey has 96% of galaxies brighter than this limit at 1 $R_e$, and 75% still brighter at 2$R_e$ if 24x1.6" cores (equivalent to outer ring of 61-core bundle) are binned. In reality the outer ring would be unlikely to be binned together, but instead an equivalent number of cores would be combined with a Voronoi binning scheme in order to retain spatial information[43]. Hector will also achieve S/N=50 over 4700-7000Å for r<22.4 mag/arcsec$^2$ in 24 binned cores in 73% of galaxies. That is sufficient to meet the stellar kinematics science case that favours predominantly galaxies at higher stellar mass. For low stellar mass or Sersic index<1.5 galaxies, S/N>10 over 4700-7000Å in 24 binned cores should be achieved for r<24.5 mag/arcsec$^2$, for 98 and 82% of all galaxies in the Hector Galaxy Survey at 1 and 2 $R_e$ respectively.

## 5.  CONCLUSIONS

Hector will be a new IFS instrument for the AAT that can, for the first time, push the size of an IFS galaxy survey to 100,000 galaxies in order to achieve new science that cannot be done with samples of a few thousand galaxies. This will be made possible with the installation of a new 3-degree diameter field top end corrector for the AAT incorporating a new mechanical design. The Hector hexabundle IFUs will be hexagonally-packed fused fibres bundles of 15-30 arcsec diameter to allow imaging of 90% of the Hector galaxy survey galaxies out to 2$R_e$. Novel robotic positioning robots, called starbugs will simultaneously align all 50-100 hexabundles across the glass field plate within a few minutes, optimising observing efficiency. A ~40m fibre cable will connect the hexabundles to custom-designed replicable spectrographs with 1.3 Å resolution from 3727-7761Å. Detailed mechanical and optical assessment of the Hector 3-degree-field corrector is underway, and the spectrographs design will be finalised in the next few months. Hector is expected to see first light in 2020. A later extension to a far-red wavelength spectrograph arm up to 1μm may follow.

### Acknowledgements

The SAMI Galaxy Survey is based on observations made at the Anglo-Australian Telescope. The Sydney-AAO Multiobject Integral-field spectrograph (SAMI) was developed jointly by the University of Sydney and the Australian Astronomical Observatory, and funded by ARC grants FF0776384 (Bland-Hawthorn) and LE130100198. The SAMI Galaxy Survey website is http:// sami-survey.org/.
Parts of this research were conducted by the Australian Research Council Centre of Excellence for All-sky Astrophysics (CAASTRO), through project number CE110001020.
Bland-Hawthorn is funded by an ARC Laureate Fellowship (FL140100278). The SAIL labs are funded through ARC Laureate and LIEF grants, external contracts and generous funding from the University of Sydney.